\documentclass[11pt]{article}
\usepackage{BSMvietnam,epsfig}
\usepackage{amssymb,amsmath}

\def\Journal#1#2#3#4{{#1} {\bf #2}, #3 (#4)}

\def\PLB{{\em Phys. Lett.}  B}

\def\PRD{{\em Phys. Rev.} D}

\def\be{\begin{equation}}
\def\ee{\end{equation}}
\def\bea{\begin{eqnarray}}
\def\eea{\end{eqnarray}}

\begin{document}
\vspace*{4cm}
\title{REVIEW ON TOP FORWARD-BACKWARD ASYMMETRY}

\author{ P. KO }

\address{School of Physics, Korean Institute for Advanced Study \\
Seoul 130-722, Korea}

\maketitle\abstracts{
The top forward-backward asymmetry (FBA) observed at the Tevatron has been 
a hot issue in particle physics for the last few years.  In this talk, 
I describe two different approaches for the top FBA at the Tevatron, one in effective 
field theory (EFT) approach and the other in explicit model for $Z'$.  Within the first approach, 
I identify a class of models which can accommodate the top FBA when new physics scale is 
very heavy. 
Axigluon or $t$-channel scalar exchanges with flavor dependent couplings can do the job. 
In the second approach, I show that the chiral couplings of $Z'$ necessarily invites 
multi Higgs doublets with $Z'$ couplings. Otherwise the top quark becomes massless, which 
is completely unphysical. Newly introduced multi Higgs doublets also contribute to the top FBA,
$\sigma_{tt}$ and the charge asymmetry at the LHC, and there are parameter regions which are
compatible with all the observations related with the top quarks.}

\section{Introduction}

The $A_{\rm FB}$ of the top quark is one of the interesting observables 
related with top quark.  Within the Standard Model (SM), 
this asymmetry vanishes at leading order in  QCD because of $C$ symmetry. 
At next-to-leading order [$O(\alpha_s^3)$], 
a nonzero $A_{\rm FB}$  can develop from the interference
between the Born amplitude and two-gluon intermediate state, 
as well as the gluon bremsstrahlung and gluon-(anti)quark scattering 
into $t \bar{t}$, with the prediction $A_{\rm FB}\sim 0.078$ \cite{Kuhn:2011ri}. 
%
The measured asymmetry has been off the SM prediction by $\sim 2 \sigma$ 
for the last few years, albeit a large experimental uncertainties. 
The averages of the lepton plus jet and the dilepton channels  in the $t\bar{t}$ rest frame
 were reported at the ICHEP2012 by both CDF and D0 Collaborations \cite{muller}: 
\begin{equation}
A_{\rm FB} \equiv \frac{N_t ( \cos\theta \geq 0) - N_{\bar{t}} 
( \cos\theta \geq 0 )}{N_t ( \cos\theta \geq 0) + N_{\bar{t}} 
( \cos\theta \geq 0 )} = 
(0.201 \pm 0.067) ({\rm CDF}) ~~~
(0.196 \pm 0.060)  ({\rm D0})  
\end{equation}
with $\theta$ being the polar angle of the top quark with 
respect to the incoming proton in the $t\bar{t}$ rest frame.
This $\sim 2\sigma$ deviation stimulated some speculations on new physics
scenarios. 
On the other hand, search for a new resonance decaying into  `
$t\bar{t}$ pair has been  carried out  at the Tevatron and the LHC.
Also there is no evidence for large top charge asymmetry at the LHC \cite{muller}. 

In this talk I describe two independent approaches for the top FBA, one in the effective 
field theory (EFT) approach, and the other within an explicit gauge model with $Z'$. 
In Sec.~2, I address the top FBA within the EFT framework, assuming that 
the  new physics scale relevant to this puzzle is very heavy and does not 
show up at the Tevatron.  
In Sec.~3, we consider the $Z'$ model, implementing it to the realistic gauge 
theory. We find that the original $Z'$ model by Jung {\it et al.}~\cite{jung} should be 
extended by introducing new Higgs doublets that couple to the $Z'$ in order to generate 
nonzero masses for the up-type quarks.  We show how this model with 
relatively light $Z'$ can satisfy all the available data as of now. 
This talk is based on a series of my works~\cite{ko-top-1-1,ko-top-1-2,ko-top-2-1,ko-top-2-2}.

\section{Effective Field Theory (EFT) Approach: the top FB asymmetry, longitudinal top polarization, 
and other observables}
\subsection{Lagrangian}
At the Tevatron, the $t\bar{t}$ production is dominated by $q\bar{q} 
\rightarrow t\bar{t}$, and it would be sufficient to consider dimension-6
four-quark operators (the so-called contact interaction terms) 
to describe the new physics effects on the $t\bar{t}$ production 
at the Tevatron  assuming that new physics scale is high enough~\cite{ko-top-1-1}:   
\begin{equation}
\mathcal{L}_6 = \frac{g_s^2}{\Lambda^2}\sum_{A,B}
\left[C^{AB}_{1q}(\bar{q}_A\gamma_\mu  q_A)(\bar{t}_B\gamma^\mu t_B)  + 
C^{AB}_{8q}(\bar{q}_A T^a\gamma_\mu q_A)(\bar{t}_B T^a\gamma^\mu  t_B)\right]
\end{equation}
where $T^a = \lambda^a /2$, $\{A,B\}=\{L,R\}$, and 
$L,R \equiv (1 \mp \gamma_5)/2$ 
with $q=(u,d)^T,(c,s)^T$.  
Using this effective lagrangian, we calculate the cross section up to 
$O(1/\Lambda^2)$, keeping only the interference term between 
the SM and new physics contributions.  

We make one comment: the chromomagnetic operators of dim-5 would 
be generated at one loop level, whereas the $q\bar{q} \rightarrow t \bar{t}$
operators can be induced at tree level. 
Therefore the chromomagnetic operators will be suppressed further by 
$g_s / ( 4 \pi )^2 \times (loop \ function)$, compared with the dim-6 
operators we consider in this talk. Therefore we will ignore chromomagnetic
operators in this talk. 

\begin{figure}
\includegraphics[width=8cm]{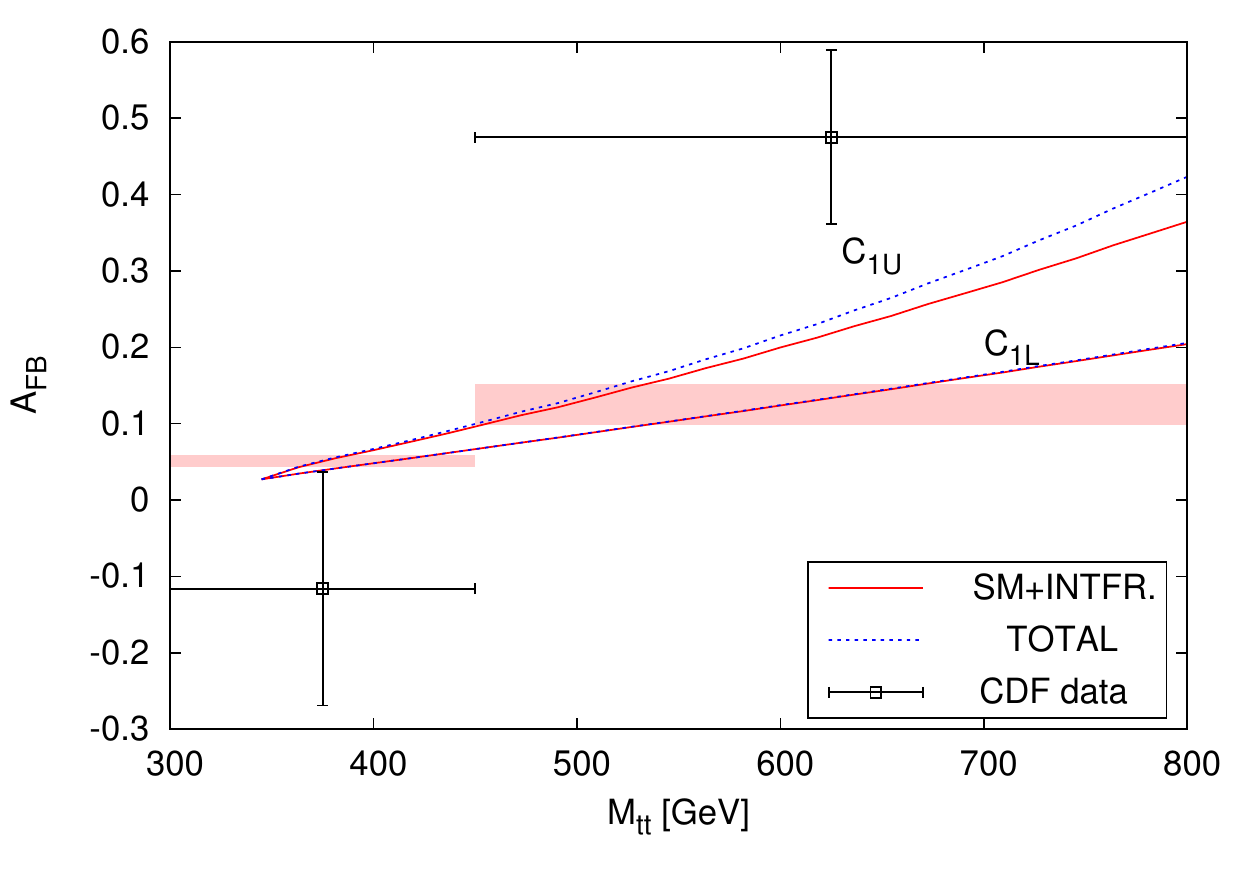}
\includegraphics[width=8cm]{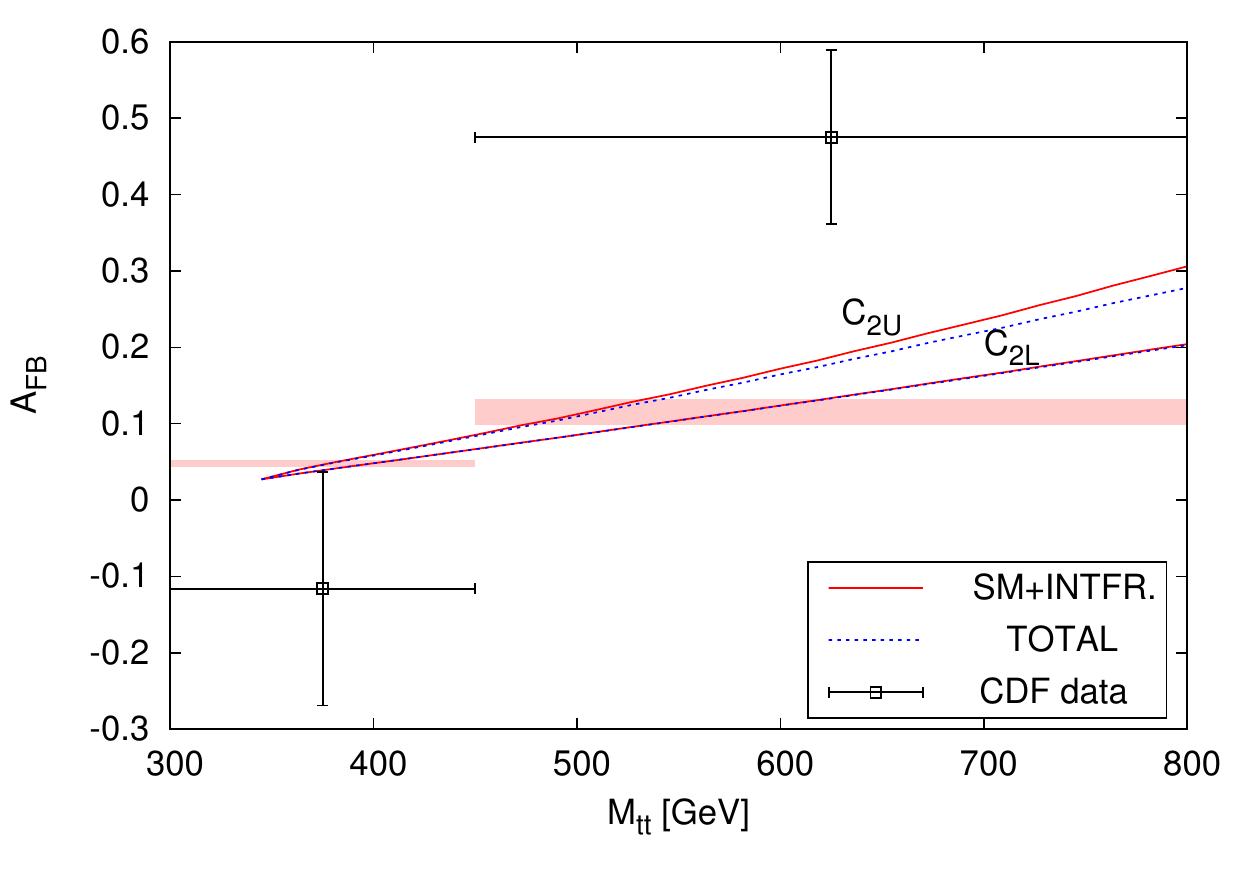} \\%
\caption{The $m_{t\bar{t}}$ distributions of the top FBA 
for (a) $C_2 = 0$ and (b) $C_1 = 0$. Blue dotted curves include the square of 
new physics amplitudes in addition to the interference term. 
}
\label{fig:1a}
\end{figure}

\subsection{Origin of FB Asymmetry}

It is straightforward to calculate the amplitude for 
$q (p_1) + \bar{q} (p_2) \rightarrow t (p_3) + \bar{t} (p_4)$ 
using the above effective lagrangian and the SM. 
The squared amplitude summed (averaged) over the final (initial) 
spins and colors is given by
\begin{eqnarray}
\overline{|{\cal M}|^2} 
& \simeq  & \frac{4\,g_s^4}{9\,\hat{s}^2} \left\{
2 m_t^2 \hat{s} \left[
1+\frac{\hat{s}}{2\Lambda^2}\,(C_1+C_2)
\right] s_{\hat\theta}^2   \right. 
\\ 
& + & \left. 
\frac{\hat{s}^2}{2}\left[ \left(1+\frac{\hat{s}}{2\Lambda^2}\,(C_1+C_2)\right)
(1+c_{\hat\theta}^2)
+\hat\beta_t\left(\frac{\hat{s}}{\Lambda^2}\,(C_1-C_2)\right)c_{\hat\theta}
\right]\right\}   \nonumber 
\label{eq:ampsq}
\end{eqnarray}
where $\hat{s} = (p_1 + p_2)^2$, $\hat\beta_t^2=1-4m_t^2/\hat{s}$,
and $s_{\hat\theta}\equiv \sin\hat\theta$ and 
$c_{\hat\theta}\equiv \cos\hat\theta$ 
with $\hat{\theta}$ being the polar
angle between the incoming quark and the outgoing top quark in the 
$t\bar{t}$ rest frame. 
And the couplings are defined as:
$C_1 \equiv C_{8q}^{LL}+C_{8q}^{RR}$ and 
$C_2 \equiv C_{8q}^{LR}+C_{8q}^{RL}$. 
Since we have kept only up to the interference terms, there are 
no contributions from  the color-singlet operators with coupling 
$C_{1q}^{AB}$. 
The term linear in $\cos\hat{\theta}$ could
generate the forward-backward asymmetry 
which is proportional to $\Delta C \equiv (C_1 - C_2)$.
Note that both light quark and top quark should have chiral couplings
to the new physics in order to generate $A_{\rm FB}$ at the tree level
(namely $\Delta C \neq 0$).  This parity violation, if large, 
could be observed in the nonzero (anti)top spin polarization \cite{ko-top-1-2}. 
We found that the Tevatron integrated top FBA can be explained for $0.15 \lesssim C_1 \lesssim 0.97$ 
or $-0.67 \lesssim  C_2 \lesssim -0.15$ at $1\sigma$ level for $\Lambda = 1$ TeV. 
Note that the negative sign of $C_2$ is preferred  at the 1 $\sigma$ level.   
In Fig.~1 (a) and (b), I show the mass dependent top FBA using the ranges of $C_1$ or 
$C_2$  determined from the integrated FBA. 
Note that the results based on the EFT is somewhat lower than the data at the
high $m_{t\bar{t}}$ region~\cite{ko-top-1-1}. 
If this disagreement does not disappear, it would imply that
the EFT approach  is not a good description for the top FBA.

\subsection{Longitudinal polarization of top quark probes chiral structures of new physics}

The top AFB is not sensitive to the detailed chiral structures of new physics, since 
the AFB depends on two couplings, $C_1 = C_{8q}^{LL} + C_{8q}^{RR}$ and 
$C_2 = C_{8q}^{LR} + C_{8q}^{RL}$, and not individual $C_{8q}^{AB}$'s.
In Ref.~\cite{ko-top-1-2}, the top longitudinal polarization was proposed as a probe of chiral
structures of new physics that would be relevant to the Tevatron top FBA. 
Note that the  longitudinal polarizations of $t$ and $\bar{t}$ vanish in QCD due to its 
parity conservation.  Any new physics for the top FBA involve chiral couplings of a new particle
to the SM quarks  and the parity will be no longer conserved when interfering with QCD.
%

Neglecting the transverse polarizations, one can derive 
\begin{equation}
\overline{|{\mathcal{M}}|^2}  =  \frac{g_s^4}{\hat{s}^2}\Bigg\{
{\cal D}_0  +
{\cal D}_1 (P_L+\bar{P}_L) 
+ {\cal D}_2 (P_L-\bar{P}_L)+
{\cal D}_3 P_L \bar{P}_L \Bigg\}\,.
\end{equation}
where $P_L$ and $\bar{P}_L$ are the longitudinal polarizations of $t$ and $\bar{t}$. 
The unpolarized coefficient ${\cal D}_0$ leads to
the total cross section $\sigma_{t\bar{t}}$ and the forward-backward
asymmetry $A_{\rm FB}$ shown in Eq.~(3).
On the other hand, the coefficient ${\cal D}_3$ gives the 
spin-spin correlations $C$ and $C_{\rm FB}$ considered 
and suggested in Ref.~\cite{ko-top-1-2}.

Note that  the other two coefficients ${\cal D}_1$ and ${\cal D}_2$
are $P$ violating. Furthermore, the coefficient ${\cal D}_1$ 
is odd  under both the CP and CP$\widetilde{\rm T}$ transformations
\footnote{The $\widetilde{\rm T}$ transformation
reverses the signs of the spins
and the three-momenta of the
asymptotic states, without interchanging initial and final
states, and the matrix element gets complex conjugated.}.
In our effective lagrangian approach, new heavy particles 
are integrated out, and there is no new strong CP-even phase, and so 
${\cal D}_1$ is zero. However, it could be nonzero when 
the heavy particle is explicitly included, and we keep
the finite decay width of the heavy particle 
together with possible CP-violating phases in its couplings to
light and top quarks.   This issue will be discussed in full 
in the future publication \cite{progress}.

\begin{figure}
\includegraphics[width=8cm]{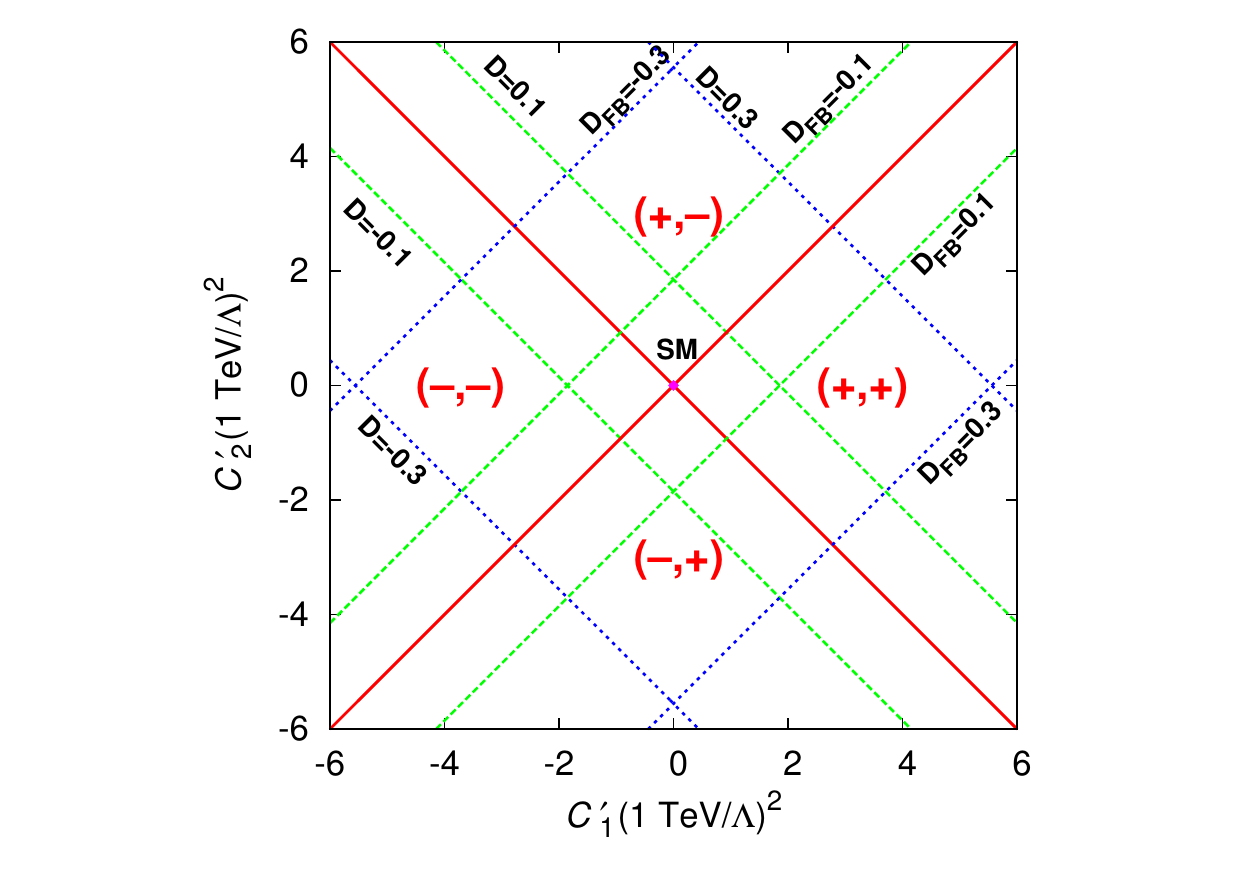}
\includegraphics[width=8cm]{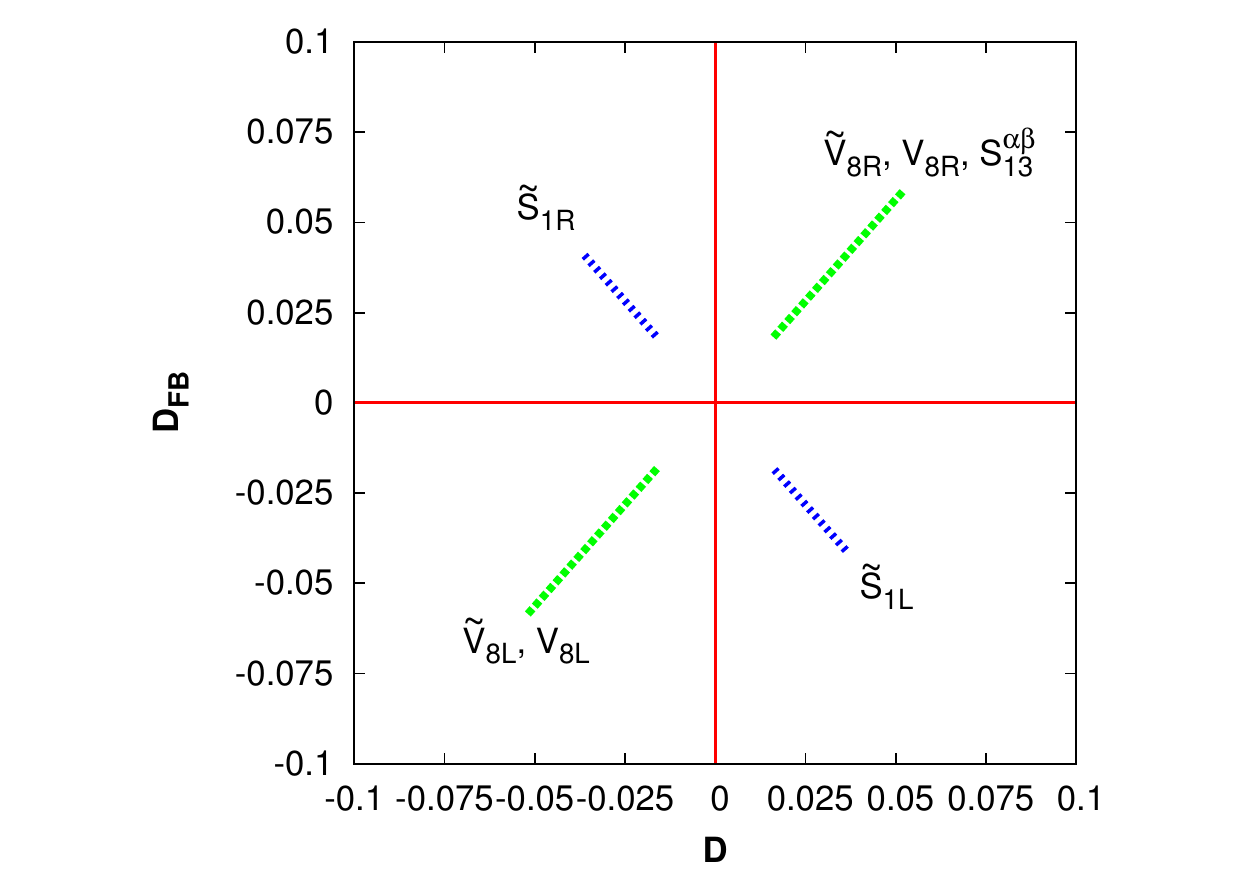}
\caption{
The $P$-violating spin correlations $D$ and
$D_{\rm FB}$ in the $(C_1^\prime,C_2^\prime)$ plane.
The signs of $(D\,,D_{\rm FB})$ are denoted.}
\label{fig:ddfb}
\end{figure}

The other $P$-violating coefficient
${\cal D}_2$ could be observable at the Tevatron, revealing
genuine features of new physics responsible for $A_{\rm FB}$.
Explicitly, we have obtained 
\begin{equation}
{\cal D}_2 \simeq \frac{\hat{s}}{9\,\Lambda^2}\left[
(C_1^\prime + C_2^\prime) \hat{\beta_t} (1+c^2_{\hat{\theta}})
 +(C_1^\prime - C_2^\prime) (5-3 \hat{\beta}_t^2) c_{\hat{\theta}}\right]
\end{equation}
with $C_1^\prime  \equiv  C_{8q}^{RR}-C_{8q}^{LL}\,, \quad 
C_2^\prime  \equiv  C_{8q}^{LR}-C_{8q}^{RL}\,.$
Therefore ${\cal D}_2$ will provide additional information
on the chiral structure of new physics in $q\bar{q} \rightarrow t\bar{t}$.
When we integrate over the polar angle $\hat{\theta}$, only the first term 
involving  $(C_1^\prime + C_2^\prime) = C_{8q}^{RR} - C_{8q}^{LL} + C_{8q}^{LR} 
- C_{8q}^{RL} $  survives. 
On the other hand, if we separate the forward and the backward top 
samples and take the difference, the orthogonal combination in the second term 
survives: $(C_1^\prime - C_2^\prime) = C_{8q}^{RR} - C_{8q}^{LL} - C_{8q}^{LR} + 
C_{8q}^{RL} \,.$
For definiteness, we consider the two new observables:
\begin{eqnarray}
D &\equiv & \frac{\sigma(t_R\bar{t}_L) - \sigma(t_L\bar{t}_R)}
{\sigma(t_R\bar{t}_R) + \sigma(t_L\bar{t}_L) +
\sigma(t_L\bar{t}_R) + \sigma(t_R\bar{t}_L)}\,,
\nonumber \\[0.1cm]
D_{\rm FB} &\equiv &
D (\cos\hat\theta \geq 0) -  D (\cos\hat\theta \leq 0)
\end{eqnarray}
which involve the sum and difference of the coefficients
$C_1^\prime$ and $C_2^\prime$, respectively.
In Fig.~\ref{fig:ddfb}, we show the $P$-violating spin correlations
$D$ and $D_{\rm FB}$ in in the $(C_1^\prime,C_2^\prime)$ plane within the 
parameter region consistent with the top FBA at the Tevatron. 
We observe that $|D|$ and $|D_{\rm FB}|$, which are zero in the SM, could be as large as $0.1$
in the region $|C_{1,2}^\prime\,(1\,{\rm TeV}/\Lambda)^2|\lesssim 1$, 
which have to be actively searched for.

\section{Beyond the EFT : the case of extra $Z^{'}$ model with flavored multi Higgs Doublets}
\subsection{Original $Z'$ model for the top FBA}
Now let us consider an explicit model, a $Z'$ model first proposed by Jung, Murayama,
Pierce and Wells ~\cite{jung}. In this model, it is assumed that there is a flavor changing couplings 
of $Z'$ to the right-handed $u$ and $t$ quarks:
\begin{equation}
{\cal L} = - g_X Z_\mu^{'} \left[  \overline{t_R} \gamma^\mu u_R + H.c. \right] .
\end{equation}
The $t$-channel exchange of $Z'$ leads to the Rutherford peak in the forward direction
and generates the desired amount of the top FBA if $Z'$ is around $150-250$ GeV and
$g_X$ is not too small. Here $Z'$ is assumed to couple only to the right-handed (RH) 
quarks in order to evade the strong bounds from the FCNS processes such as 
$K^0 - \overline{K^0}$,
$B^0_{d(s)} - \overline{B_{d(s)^0}}$ mixings and $B \rightarrow X_s \gamma$. And such a 
light $Z^{'}$ should be leptophobic in order to avoid the strong bounds from the Drell-Yan 
processes. Therefore the original $Z'$ model is chiral, leptophobic and flavor non universal. 
It would be nontrivial to construct a realistic gauge theory which satisfies these conditions.
Also the original $Z'$ model was excluded by the same sign top pair productions, because 
$Z'$ exchange can contribute to $ u u \rightarrow t t$. 

\subsection{$U(1)'$ models with flavored multi Higgs doublets by Ko, Omura and Yu}

In Ref.s~\cite{ko-top-2-1,ko-top-2-2}, realistic models for such peculiar $Z'$ have been constructed, 
and were shown to be less constrained by the same sign top pair production. 
In Ref.s~\cite{ko-top-2-1,ko-top-2-2},  $Z'$ is assumed to be a gauge boson of a new $U(1)'$ 
local gauge symmetry, 
under which the RH quarks are charged non universally, whereas other quarks are charged
universally or completely neutral. 
Then one cannot write renormalizable Yukawa couplings to the RH up-type quarks,
since $\overline{Q_L} \widetilde{H} u_R$ is not gauge invariant, where $H$ is the SM Higgs doublet
which is $U(1)'$ singlet and generates masses for the down-type quarks and charged leptons. 
We have to introduce new Higgs doublets $H_i$ which are charged under the new $U(1)'$.
We can also introduce a singlet scalar which carries $U(1)'$ charge and breaks $U(1)'$ 
spontaneously. However it is not mandatory to introduce $\Phi$, since the newly introduced 
$U(1)'$-charged Higgs doublets $H_i$ also break $U(1)'$ and generate $Z'$ mass too.
The offshot of the chiral $Z'$ model is that one has to extend the Higgs sector, whether
$Z'$ couplings are flavor universal or not, as long as the $Z'$ couplings are chiral.
\footnote{The model has gauge anomaly, and one has to add some fermions. 
This issue is described in Ref.~\cite{ko-top-2-1} in  detail.}
 The same 
arguments apply to other models with  new spin-1 objects that have chiral couplings to the 
SM fermions, including the axigluon, extra $W'$, $SU(2)$ $W_I$, $Z_I$ and $SU(3)_R$ flavor
gauge bosons.  

Introducing $U(1)'$ flavored Higgs doublets is very important because they generates nonzero 
top mass.  They also play an important role in top FBA phenomenology.  
For example the Yukawa couplings of the neutral scalar bosons $h,H,a$ have flavor changing
couplings to the up-type quarks because of the flavor non universal nature of $Z'$ interaction~
\cite{ko-top-2-1,ko-top-2-2}:
\begin{eqnarray} 
Y^h_{tu} & = & \frac{ 2m_t (g^u_R)_{ut} }{v \sin( 2 \beta)} 
\sin (\alpha-\beta) \cos \alpha_{\Phi} \ ,  
\\
Y^H_{tu} & = & - \frac{ 2m_t (g^u_R)_{ut} }{v \sin( 2 \beta)} 
\cos (\alpha-\beta) \cos \alpha_{\Phi} \ ,
\\
Y^a_{tu} & = & \frac{ 2m_t (g^u_R)_{ut} }{v \sin( 2 \beta)} \ .
\end{eqnarray} 

These Yukawa couplings are not present in the Type-II 2HDM, for example.  
Our models  proposed in Ref.s~\cite{ko-top-2-1,ko-top-2-2} are good examples of non minimal 
flavor violating multi  Higgs doublet  models. 
The top FBA and the same sign top pair productions are generated not only by the 
$t$-channel $Z'$ exchange, but also by the $t$-channel exchange of neutral Higgs scalars,  
and the strong constraint on the original $Z'$ model  from the same sign top pair production 
can be relaxed by a significant amount. 
The results with a few low lying (pseudo)scalar bosons as well as $Z'$ are shown in Fig.~3. 
Note that the strong bound from the same sign top pair production can be evaded because of 
the $t-$channel exchanges of $h$ and $a$.  And the $m_{t\bar{t}}$ distribution becomes closer
to the SM case in the presence of $h$ and $a$ contributions. (See Ref.~\cite{ko-top-2-2} for
more detailed discussions on top charge asymmetry at the LHC and Ref.~\cite{progress2} for
the $B$ physics constraints on these models.)

\begin{figure}[!t]
\begin{center}
\includegraphics[width=6cm]{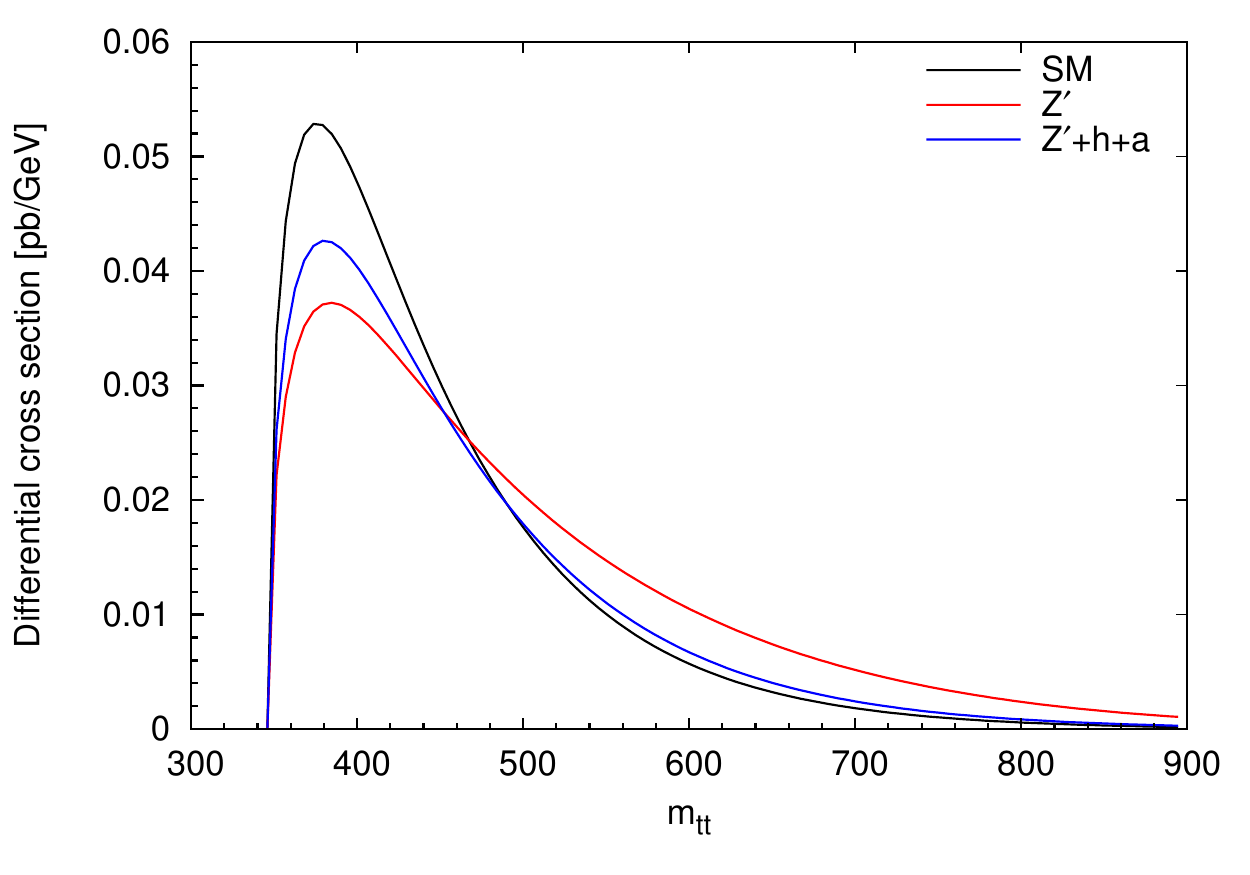}
\includegraphics[width=6cm]{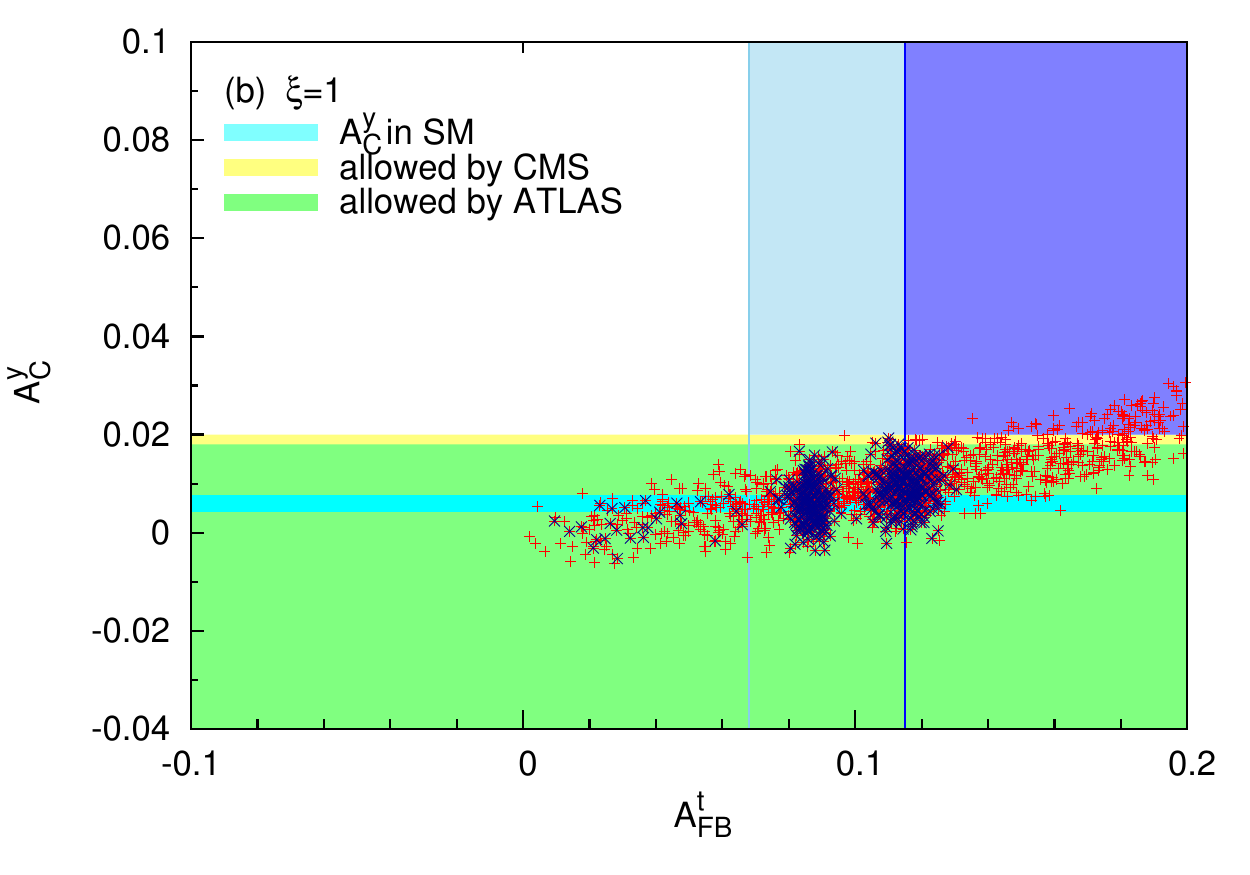}
\includegraphics[width=6cm]{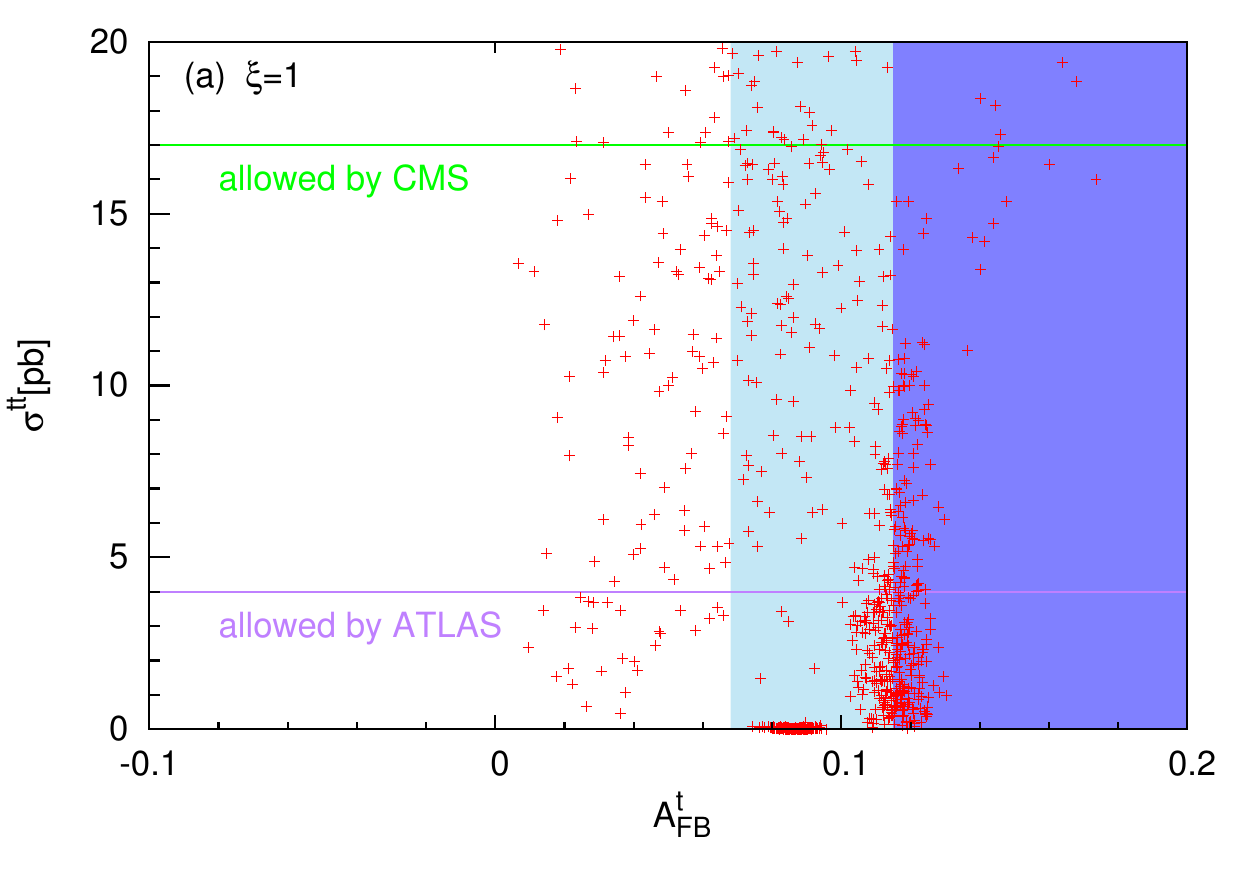}
\includegraphics[width=6cm]{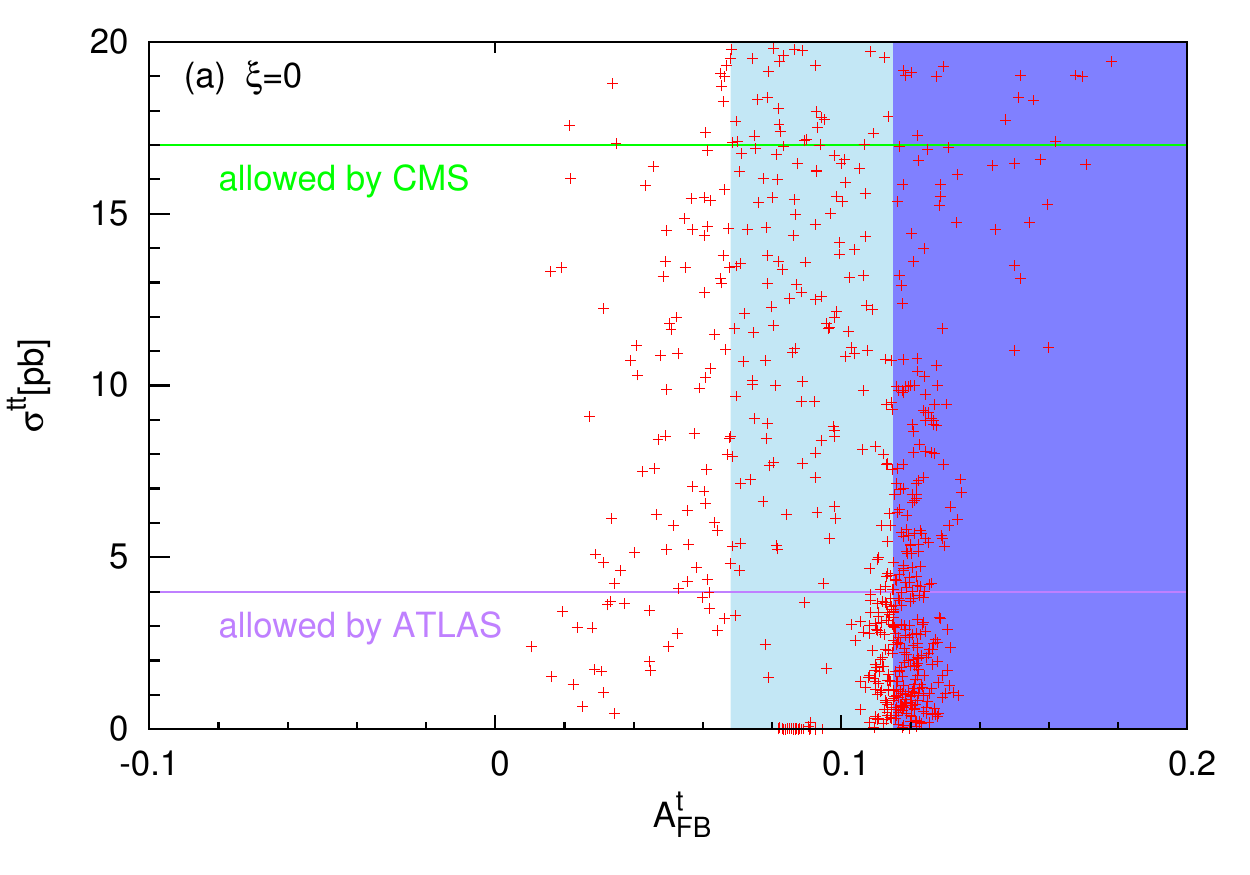}
\caption{\label{fig:higgs1}%
The scattered plots for (a) the $m_{tbar{t}}$ distributions for the SM, SM+$Z'$ and
SM + $Z' , h + a$, (b) $A_\textrm{FB}^t$ at the Tevatron and
$\sigma^{tt}$ at the LHC for $\xi = 1$, and  
(c)  and (d)  $A_\textrm{FB}^t$ at the Tevatron and the charge asymmetry 
$A_C^y$ at the LHC for $\xi=1$ and $\xi=0$, respectively.
}
\end{center}
\end{figure}

\section{Conclusions}

In this talk, I discussed two independent approaches for the top FB asymmetry observed 
at the Tevatron in the framework of new physics models beyond the SM.  
Whatever the new physics solutions may be, there are some common features:
we need new interactions with chiral couplings to the top and the light quarks, 
and nontrivial flavor structures in the quark sector, which is not easy to
achieve in any realistic models because of the strong constraints from $K$ and $B$ 
meson systems. 
If the new physics involves chiral couplings of new spin-1 vector boson, one 
has to extend the Higgs sector too, by including new Higgs doublets that couple to
the new spin-1 vector boson. Otherwise, the model cannot give the up-type quark masses. 
All the top-related observables we are interested in, such as top FB asymmetry 
at the Tevatron, the same sign top pair production cross section and top charge  asymmetry 
at the LHC, are crucially dependent on the extended Higgs sector as well as the spin-1
objects with chiral couplings, and 
it is meaningless to discuss the phenomenology without the new Higgs doublets.

\section*{Acknowledgments}
The author thanks the organizers for the invitation and the nice organization. 
He is also grateful to Dong Won Jung, Jae Sik Lee, S.-h. Nam, Yuji Omura and 
Chaehyun Yu for fruitful collaborations on the topics discussed in this talk.
This work is supported by NRF Research Grant  2012R1A2A1A01006053, 
and by SRC program of NRF Grant No. 20120001176 through Korea Neutrino Research 
Center at Seoul National University.

\end{document}